\documentclass[aps,prapplied,reprint,groupedaddress,amsmath,amssymb,floatfix,longbibliography]{revtex4-2}
\ifdefined\pdfminorversion
\fi
\usepackage{graphicx}
\usepackage{hyperref}
\usepackage{microtype}
\hypersetup{hidelinks}
\newcommand{\id}{\mathbb I}
\newcommand{\Tr}{\operatorname{Tr}}
\newcommand{\sinc}{\operatorname{sinc}}
\newcommand{\dd}{\mathrm{d}}
\newcommand{\Del}{\Delta\omega}
\newcommand{\KFD}{K^{\mathrm{FD}}}
\newcommand{\norm}[1]{\left\lVert#1\right\rVert}
\newcommand{\ket}[1]{\lvert #1\rangle}
\newcommand{\bra}[1]{\langle #1\rvert}
\newcommand{\braket}[2]{\langle #1\vert #2\rangle}
\newcommand{\ii}{\mathrm i}
\begin{document}
\title{Simple broadband signal detection at the fundamental limit}
\author{Anthony M. Polloreno}
\email{ampolloreno@gmail.com}
\author{Graeme Smith}
\affiliation{Department of Applied Mathematics and Institute for Quantum Computing, University of Waterloo, 200 University Avenue West, Waterloo, Ontario N2L 3G1, Canada}
\date{November 30, 2026}
\begin{abstract}
Broadband detection of a weak oscillatory field with unknown carrier frequency underlies magnetometry, axion searches and gravitational-wave sensing. We show that the Grover-like integration-time lower bound for this task is a geometric corollary of an upper bound on the integrated quantum Fisher information and present an all-analog protocol with near-optimal scaling.
\end{abstract}
\maketitle

\section{Introduction}
Quantum sensors convert weak classical fields into measurable changes of a quantum state \cite{degen,brown}. For oscillating signals, coherent control enables strong narrowband selectivity, but in many settings the carrier frequency is not known a priori and may lie anywhere within a wide band. The resulting bandwidth--sensitivity tradeoff is a recurring bottleneck in AC magnetometry. Applications of frequency-selective measurements include noise spectroscopy \cite{bylander}, searches for axion-like dark matter \cite{graham}, and gravitational-wave detection \cite{ligo}.

A recent line of work on quantum-algorithmic sensing emphasizes that broadband detection resembles an unstructured search problem and derives a sensing-time lower bound by analogy with Grover's algorithm \cite{grover,allen}. Recent qubit--cavity experiments have also implemented Grover-based searches for signals of unknown frequency \cite{xu,ouellet}. This raises a natural question: \emph{is saturating the lower bound on broadband sensing inherently computational (requiring a quantum computer), or is it fundamentally metrological (achievable by coherent control and measurement alone)?} We argue for the latter, within a coherent-control model. We connect broadband detection directly to the integrated quantum Fisher information (IQFI) \cite{polloreno}, identify the finite-amplitude statistical task that makes its geometry relevant, and give a multiresonant analog construction. We compare its simulated detection times with the predicted scaling.

For a known finite set of candidate frequencies, Chen \emph{et al.} attain the discrete Grover--Heisenberg limit for detecting a single tone with one qubit \cite{chen_discrete}. Liu \emph{et al.} use a single qubit with $\pi$ pulses for detection and search when a signal can occupy several of the candidate frequencies \cite{liu_discrete}. Our construction covers a continuous band by alternating frequency-selective phase marks with a bus-mediated mixing operation, without an explicit quantum signal processing digitization step. For stationary marks we prove a uniform detection bound by alternating the sign. For a continuously running tone, the same bus reflection can read out the phase of the elementary sensing unit of Allen \emph{et al.}~\cite{allen} directly. This second construction attains the broadband time bound up to a square root of a logarithm, with a separate low-frequency cost when the band approaches zero. In the title, ``simple'' refers to this mark--mix control sequence. The time bound counts interrogation, and the receiver also requires an auxiliary frequency bank, state preparation, a phase reference, and calibration.

\section{Broadband AC detection problem}
We formalize broadband detection (as in Ref.~\cite{allen}) as a promise problem $\mathrm{AC}[B_{\min},\Del]$: given an oscillatory field
\begin{equation}
B(t)=B\cos(\omega t+\phi),\qquad \omega\in\Del=[\omega_{\min},\omega_{\max}],
\label{eq:field}
\end{equation}
with unknown amplitude $B$ and phase $\phi$, decide whether $B=0$ or $B\ge B_{\min}>0$. We take $0<\omega_{\min}<\omega_{\max}$ and write $|\Del|=\omega_{\max}-\omega_{\min}$ for the bandwidth. At zero carrier and phase $\pi/2$, the alternative field would vanish identically. Both error probabilities must be at most a fixed $\epsilon<1/2$, uniformly over the promised alternatives. For $m$ field-coupled sensors we focus on the genuinely broadband regime
\begin{equation}
\beta_{\min}=mB_{\min},\qquad r=|\Del|/\beta_{\min}\gg1.
\label{eq:scales}
\end{equation}
The converse below makes no high-carrier-frequency assumption. For the receiver, whose amplitude scan must be finite, we additionally specify $B\in[B_{\min},B_{\max}]$ with $mB_{\max}\le |\Del|$.

\begin{figure*}[t]
\centering\includegraphics[width=0.94\textwidth]{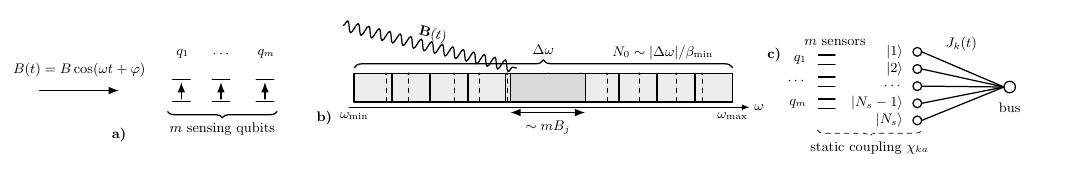}
\caption{Analog coherent broadband detection. (a) A Greenberger--Horne--Zeilinger (GHZ) probe of $m$ sensors $q_a$ supplies the effective signal scale $\beta=mB$ in the commuting mark model. (b) A bucket bank covers the frequency band with linewidths of order $mB_j$. (c) A candidate circuit-QED layout uses bucket-dependent sensor shifts $\chi_{ka}$ and controlled bucket--bus exchange amplitudes $J_k(t)$. Realizing the effective mark and a sensor-independent mixer requires the modulation, phase-reference, and frame-control operations discussed in Appendix~\ref{app:physics}.}
\label{fig:schematic}
\end{figure*}

Our setup is shown in Fig.~\ref{fig:schematic}. Let $\rho(B,\omega,\phi;T)$ be the state after a total interrogation time $T$ under a controlled Hamiltonian
\begin{equation}
H(t)=G(t)+B\cos(\omega t+\phi)Z_{\rm tot},\quad
Z_{\rm tot}=\id_{\rm aux}\otimes\sum_{\ell=1}^{m}Z^{(\ell)},
\label{eq:hamiltonian}
\end{equation}
where $Z^{(\ell)}$ is the single-sensor signal generator on sensing subsystem $\ell$, with $\norm{Z^{(\ell)}}\le1$. The initial state and coherent control $G(t)$ may depend on the promised band and threshold but not on the unknown $B,\omega,\phi$. There may be any finite number of auxiliary modes; they do not contribute to $m$. We assume a well-defined unitary propagator and a fixed maximum interrogation duration, with parameter-independent feedback coherently deferred if necessary.

\section{Integrated quantum Fisher information and detection}
Let $J(B\mid\omega,T)$ denote the quantum Fisher information (QFI) of the final state with respect to $B$ at fixed $\omega,\phi$, keeping the control protocol fixed. The band-integrated sensitivity is the IQFI \cite{polloreno},
\begin{equation}
K(B,T)=\int_{\Del}\dd\omega\,J(B\mid\omega,T).
\label{eq:iqfi}
\end{equation}
A related frequency-integrated sensitivity bound applies to bosonic noise sensing \cite{shi}. The phase is suppressed in Eq.~\eqref{eq:iqfi}; every statement used below either holds at each phase or specifies an average. An important distinction in IQFI is between the perturbative regime $mBT\lesssim1$, where integrated sensitivity can scale linearly with time, and the nonperturbative regime, where quadratic growth is possible \cite{polloreno}. Broadband detection, however, is not the same experiment as locally estimating the amplitude at $B=0$.

\subsection{From distinguishability to an IQFI witness}
The QFI induces the Bures geometry of quantum states \cite{braunstein,uhlmann}. Fix $(\omega,\phi,T)$ and consider the one-parameter family $\rho_{B'}\equiv\rho(B',\omega,\phi;T)$, $B'\in[0,B]$, which traces out a curve on the state manifold. Using the squared Uhlmann fidelity $F(\rho,\sigma)=(\Tr\sqrt{\sqrt\rho\,\sigma\sqrt\rho})^2$, define
\begin{equation}
p_{\rm det}(B,\omega,T)=1-F(\rho_0,\rho_B),\quad
\theta_\omega=\arcsin\sqrt{p_{\rm det}}.
\label{eq:fidelity}
\end{equation}
For a pure null state, $p_{\rm det}$ is exactly the rejection probability of the projector orthogonal to that state. For mixed null states it is a geometric witness, not generally one specified measurement probability. The Bures angle $\theta_\omega$ gives the geodesic distance, while the QFI supplies the length of the amplitude path. Therefore
\begin{equation}
\theta_\omega(B,T)\le\frac12\int_0^B\dd B'\sqrt{J(B'\mid\omega,T)}.
\label{eq:path}
\end{equation}
By Cauchy--Schwarz, we obtain a finite-displacement witness
\begin{equation}
\begin{split}
\KFD(B,T)&:=\frac4{B^2}\int_{\Del}\dd\omega\,\theta_\omega(B,T)^2\\
&\le\frac1B\int_0^B\dd B'\,K(B',T).
\end{split}
\label{eq:geometry}
\end{equation}
The physical test compares the null to a finite signal-induced displacement, and local QFI measures the length of a path between them. In particular, a nonzero $K(0,T)$ is fully consistent with zero probability of rejecting the calibrated null.

The perturbative interval is important in applying Eq.~\eqref{eq:geometry}. Suppose, for a specified control family, the following IQFI estimates hold with constants uniform in the path amplitude and all promised parameters:
\begin{equation}
K(B',T)\le\begin{cases}
C_p m^2T,&mB'T\le1,\\
C_n m^3B'T^2,&mB'T\ge1.
\end{cases}
\label{eq:conditional}
\end{equation}
For an endpoint $mBT\ge1$, both branches obey the same envelope $K(B',T)\le C m^3BT^2$ throughout $0\le B'\le B$, with $C=\max\{C_p,C_n\}$. The $B$ on the right is the fixed alternative, not the running amplitude $B'$. Equivalently, splitting Eq.~\eqref{eq:geometry} at $B_c=(mT)^{-1}$ gives
\begin{equation}
\begin{split}
\KFD(B,T)&\le\frac{C_pm}{B}+\frac{C_n}{2}m^3BT^2\\
&=O(m^3BT^2),\quad mBT\ge1.
\end{split}
\label{eq:crossover}
\end{equation}
At large $mBT$, the first term in the upper bound is $O((mBT)^{-2})$ times the second. This recovers the IQFI bound on finite displacement without extending a bound proportional to $B'$ down to zero. Equation~\eqref{eq:conditional} is a uniform hypothesis on a control family, not a universal local-IQFI theorem. The next section proves the detection limit directly.

\begin{figure*}[t]
\centering\includegraphics[width=0.85\textwidth]{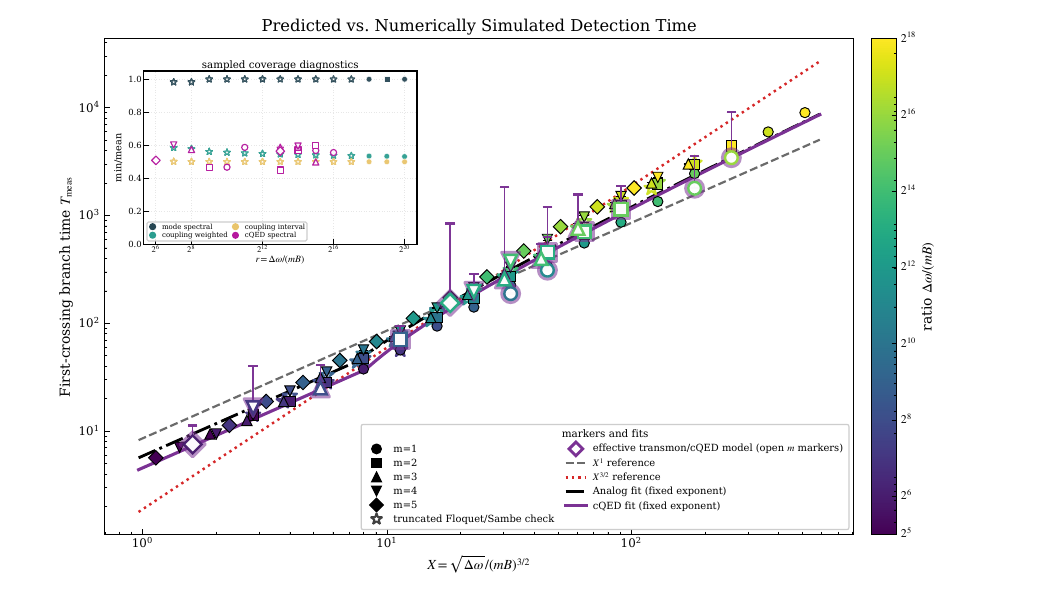}
\caption{First-crossing branch time $T_{\rm meas}=3\pi Q/(mB)$ versus $X=\sqrt{|\Del|}/(mB)^{3/2}$ for the fixed-sign receiver. Filled markers show 70 GHZ code-space groups, open stars 34 truncated Floquet/Sambe checks, and open $m$ markers 17 selected effective transmon/cQED groups. Black and purple curves show the fitted $AX\sqrt{q_{\log}(r)}$ guides. The $X$ and $X^{3/2}$ lines are reference powers. Code-space whiskers span five carrier grids. Transmon whiskers extend to the sampled maximum. The inset gives sampled spectral and bus-weight ratios. Four cQED inset points use different banks from their timing markers (Appendix~\ref{app:numerics}). Running both quadratures sequentially takes $2T_{\rm meas}$.}
\label{fig:figure2}
\end{figure*}

\subsection{General coherent control and the sensing-time limit}
We treat the time-dependent control Hamiltonian directly. In the interaction picture of $G(t)$, the signal couples through a bounded operator $A(t)$ with $\norm{A(t)}\le1$. A Fourier short-time estimate and an exact telescoping argument (Appendix~\ref{app:converse}) yield
\begin{equation}
\KFD(B,T)\le C_{\rm FD}m^2T(1+mBT),\quad C_{\rm FD}=2\pi^3e^2.
\label{eq:general}
\end{equation}
This bound holds for arbitrary parameter-independent coherent control with a well-defined propagator, independently of auxiliary dimension, control switching rate, and bandwidth. It is a statement about finite displacement; it does not imply the nonperturbative branch of Eq.~\eqref{eq:conditional} for arbitrary controls.

Any detector whose two errors are at most $\epsilon$ must produce states of trace distance at least $1-2\epsilon$. The fidelity inequalities \cite{fuchs} therefore imply $\theta_\omega(B_{\min},T)\ge\theta_\epsilon:=\arcsin(1-2\epsilon)$ throughout the band, and hence
\begin{equation}
\KFD(B_{\min},T)\ge\frac{4\theta_\epsilon^2}{B_{\min}^2}|\Del|.
\label{eq:mustdetect}
\end{equation}
Set $y=\beta_{\min}T$. Equations~\eqref{eq:general} and \eqref{eq:mustdetect} give $4\theta_\epsilon^2r\le C_{\rm FD}y(1+y)$. In a broad band, $y$ cannot remain perturbative, and
\begin{equation}
T=\Omega\!\left(\frac{\sqrt{|\Del|}}{(mB_{\min})^{3/2}}\right),\qquad r\gg1.
\label{eq:limit}
\end{equation}
This recovers the sensing-time scaling of Ref.~\cite{allen} as a direct consequence of state distinguishability. The signal-generator norm gives the factor $m$, so the converse does not rely on GHZ inputs. In the language of IQFI, it is the transition from small-amplitude geometry to finite displacement---rather than local amplitude estimation at the null---that sets the broadband cost.

\paragraph*{Single-frequency inference via spectral flatness.} In a sensing instance there is a fixed unknown $\omega^\star$. To relate single-frequency data to the band-integrated witness, one can independently certify that the response has no deep spectral holes. Write $S_\omega=\theta_\omega^2$ and $\bar s=|\Del|^{-1}\int_{\Del}S_\omega\dd\omega$. A full-response certificate
\begin{equation}
\sup_{\omega\in\Del}\left|\frac{S_\omega}{\bar s}-1\right|\le\varepsilon_T<1
\label{eq:flat}
\end{equation}
would bound the integral by $|\Del|S_\omega/(1+\varepsilon_T)$ and $|\Del|S_\omega/(1-\varepsilon_T)$. This is a conditional inference tool, not a certificate supplied by local bucket counts or the finite samples in Fig.~\ref{fig:figure2}. Appendix~\ref{app:cert} gives sufficient continuous-parameter checks of the full measurement.

\section{Analog bucket protocol}
Set $\beta_j=mB_j=2^j\beta_{\min}$ for $j=0,\ldots,J=\lfloor\log_2(B_{\max}/B_{\min})\rfloor$. The receiver uses $m$ field-coupled sensing qubits, an auxiliary bank of $N_s$ bucket modes $\ket{k}$ and one auxiliary bus mode $\ket c$. The buckets and bus are orthogonal states of a single-excitation auxiliary sector, rather than additional field-coupled sensors. With $P_k=\ket{k}\bra{k}$, put
\begin{equation}
A_{\rm aux}=\sum_{k=1}^{N_s}\nu_kP_k,\qquad
\ket u=\sum_{k=1}^{N_s}g_k\ket k,\quad\sum_k|g_k|^2=1.
\label{eq:bank}
\end{equation}
A regular bank with $N_s=\lceil8r\rceil$ will suffice for the uniform ideal-model guarantee. Alternatively, $N_s=O[r\log(r/\eta)]$ iid uniform samples give coverage with design failure at most $\eta$ (Appendix~\ref{app:bank}). The star mixer addresses $\ket u$ and a single bus, not each marked frequency individually.

Because the field phase is unknown, we consider quadratures $\chi=x,y$, with $\phi_y=\phi_x+\pi/2$. The physical Pauli spin operators are $J_\chi=\tfrac12\sum_{a=1}^m\sigma_\chi^{(a)}$. On the GHZ code spanned by $\ket{R_\chi^\pm}=\ket{\pm\chi}^{\otimes m}$, $J_\chi=(m/2)Z_\chi^{\rm code}$, so the input $\ket{\mathrm{GHZ}_\chi}=(\ket{R_\chi^+}+\ket{R_\chi^-})/\sqrt2$ acquires the expected $m$-fold phase. This is an exact reduction for the commuting scalar-mark model specified next; a generic transverse physical drive does not preserve this two-dimensional code (Appendix~\ref{app:physics}).

Choose a bounded mark envelope $|f_M|\le1$ and $\tau_{M,j}=c_M/\beta_j$. During a mark window starting at global time $t_n$, let the ideal bucket-dependent Hamiltonian be
\begin{align}
H^{\rm eff}_{k,\chi,n}(t_n+t)
&= B f_M(t/\tau_{M,j})J_\chi\nonumber\\
&\quad\times\cos[\delta_k(t_n+t)+\phi-\phi_\chi],\label{eq:markham}\\
\delta_k&=\omega-\nu_k,\quad 0\le t\le\tau_{M,j}.
\end{align}
Integrating this commuting Hamiltonian gives a diagonal phase mark,
\begin{align}
M_{j,\chi,n}&=\sum_k P_k\otimes e^{-\ii\theta_{k,\chi,n}Z_\chi^{\rm code}/2}+\ket c\bra c\otimes\id,\nonumber\\
\theta_{k,\chi,n}&=mB\tau_{M,j}\operatorname{Re}\!\left[e^{\ii(\phi-\phi_\chi+\delta_kt_n)}A_j(\delta_k)\right],\nonumber\\
A_j(\delta)&=\int_0^1\dd u\,f_M(u)e^{\ii\delta\tau_{M,j}u}.
\label{eq:mark}
\end{align}
At a common cycle start, the quadratures satisfy $\theta_{k,x,n}^2+\theta_{k,y,n}^2=(mB\tau_{M,j})^2|A_j(\delta_k)|^2$. For $mB\in[\beta_j,2\beta_j)$, at least one quadrature therefore gives an order-one phase wherever $|A_j(\delta_k)|$ is bounded below by a positive constant. The rectangular envelope used below has such a passband of width $\Theta(\beta_j)$. The final detection probability also depends on the mixer and readout.

After the mark, a bus window applies
\begin{equation}
\begin{split}
G_B(t)&=\lambda_j(t)(\ket u\bra c+\ket c\bra u)\otimes\id,\\
S_j&=e^{-\ii\vartheta_{B,j}(\ket u\bra c+\ket c\bra u)\otimes\id},
\end{split}
\label{eq:bus}
\end{equation}
where $\vartheta_{B,j}=\int\lambda_j(t)\dd t$. For $\vartheta_{B,j}=\pi$ and an initially empty bus, its action on the buckets is the reflection $R=\id-2\ket u\bra u$. With $|\lambda_j|=\Theta(\beta_j)$, a mark--bus cycle costs $\Theta(\beta_j^{-1})$ interrogation time. A sequence of $Q$ cycles is the ordered product $S_{j,Q-1}M_{j,\chi,Q-1}\cdots S_{j,0}M_{j,\chi,0}$; it becomes $(S_jM_{j,\chi})^Q$ only for a stationary mark. Equation~\eqref{eq:mark} shows why that stationarity cannot simply be assumed in a continuously running laboratory signal.

For a frequency-matched bucket set, the local bus weight is
\begin{equation}
W_j(\omega)=\sum_{|\omega-\nu_k|\le\beta_j}|g_k|^2
=\Theta\!\left(\min\left\{\frac{\beta_j}{|\Del|},1\right\}\right),
\label{eq:coverage}
\end{equation}
with uniform constants, including at band edges, for equal weights $|g_k|^2=1/N_s$ and, for a random bank, on the coverage event of Appendix~\ref{app:bank}. Spectral coverage and bus weight are necessary ingredients of coherent amplification, but the amplitude and readout of the full mark--mixer sequence matter. The following bound uses alternating mark signs.

\subsection{Coherent amplification with alternating marks}
Take the centered rectangular mark with $\beta_j\tau_{M,j}=2\pi$ and the $\pi$ bus mixer, as in the simulations. In an ideal stationary, reversible-mark model, suppose that a $\pi$ shift of a controllable reference reverses exactly the preceding mark. Let $D=\sum_k e^{-\ii\eta_k}P_k$ denote that mark in one GHZ eigenbranch. Alternating signs gives a pair of cycles
\begin{equation}
W=RD^\dagger RD=R_uR_{D^\dagger u},\qquad
\cos\vartheta=|\braket u{D u}|,
\label{eq:reflections}
\end{equation}
where $R_v=\id-2\ket v\bra v$. This is the product of two reflections, as in quantum amplitude amplification \cite{brassard}, so after $q$ pairs the calibrated null-rejection probability is exactly $\sin^2(2q\vartheta)$, including both GHZ eigenbranches.

For every promised amplitude one dyadic window obeys $1\le B/B_j<2$. Draw a reference angle uniformly once for that branch, keeping it fixed over all pairs. With a constant probability the near and far buckets acquire phases separated by a fixed amount; on a regular bank this gives $\sin^2\vartheta\ge a/r_j$, where $r_j=|\Del|/\beta_j$ and $a>0$ is an absolute constant. Randomizing $q$ over $1,\ldots,O(\sqrt{r_j})$, as in randomized quantum search \cite{boyer}, gives an order-one rejection probability conditional on that reference event. For bounded $r_j$, a single-mark test supplies a constant rejection probability. Appendix~\ref{app:amplification} proves that the complete scan therefore has
\begin{equation}
p_{\rm null}=0,\qquad \inf_{B,\omega,\phi}p_{\rm signal}\ge p_1>0
\label{eq:gap}
\end{equation}
for an absolute $p_1$, including at band edges, with $N_s=O(r)$ regular modes. The bound is conservative, but independent of $r$, the carrier, the amplitude, the phase and $m$.

For a running tone, the detuning advances the relative phase between rectangular pulses, so a later $\pi$ reference shift need not invert the preceding mark (Appendix~\ref{app:physics}). Appendix~\ref{app:continuous} uses a smooth pulse that implements the same phase mark at every start time to controlled accuracy.

\section{Testing and sensing time}
Initialize $\ket{\psi_{\rm in,\chi}}=\ket u\otimes\ket{\mathrm{GHZ}_\chi}$. The family of experiments scans the $J+1$ amplitude windows and, if both are used, the two quadratures, on independently prepared probes. For each branch measure rejection by its calibrated null projector. Let $p_b$ be its marginal rejection probability and $p_{\rm fam}$ the probability that any branch rejects. Always $\max_b p_b\le p_{\rm fam}\le\sum_bp_b$; the product formula additionally requires independent branch outcomes. A device-level certificate requires
\begin{equation}
p_{\rm fam}(0)\le p_0,\qquad
\inf_{B,\omega,\phi}p_{\rm fam}(B,\omega,\phi)\ge p_1>p_0.
\label{eq:family}
\end{equation}
The ideal alternating-sign receiver supplies Eq.~\eqref{eq:family} with $p_0=0$. With implementation noise, both the aggregated null budget and the signal contrast must be calibrated; Appendix~\ref{app:cert} gives one sufficient continuous-parameter certificate.

For an amplitude window $j$, at most $Q_j=O(\sqrt{r_j})$ mark pairs of duration $\Theta(\beta_j^{-1})$ are needed. The full dyadic scan costs
\begin{equation}
T_{\rm coh}=O\!\left(\sum_{j=0}^{J}\frac{\sqrt{|\Del|}}{\beta_j^{3/2}}\right)
=O\!\left(\frac{\sqrt{|\Del|}}{\beta_{\min}^{3/2}}\right).
\label{eq:sum}
\end{equation}
This includes the amplitude scan, and the two quadratures contribute only a constant factor. Repeating whole trials on fresh probes $O[\log(1/\delta)]$ times gives fixed target error $\delta$ for the ideal null-projector test; a general independently repeated family with contrast $\Delta p=p_1-p_0$ admits a Hoeffding bound \cite{hoeffding} with $O[\Delta p^{-2}\log(1/\delta)]$ trials. Thus the stationary ideal receiver meets the lower-bound interrogation-time scaling of Eq.~\eqref{eq:limit}. An iid bank uses $O[r\log(r/\eta)]$ auxiliary modes and the same scaling on its good-bank event, with a separate failure probability $\eta$.

\paragraph*{Achievability for a continuously running tone.}
For the field in Eq.~\eqref{eq:hamiltonian}, take Pauli sensors and allow fast, bucket-controlled rotations of their logical qubit. We replace the rectangular mark by the smooth sensing pulse of Ref.~\cite{allen}. Its return phase in one frequency bin is separated from those in the other bins, uniformly in the field's initial phase. A logical spin flip on either side of a later pulse reverses its effective phase. In the successful bank of $N$ modes this separation gives $|\sin(2\vartheta)|=\Omega(N^{-1/2})$. Averaging $\sin^2(2q\vartheta)$ over $O(\sqrt N)$ possible depths $q$ gives a constant rejection probability. The accumulated pulse error at those depths is $o(1)$, uniformly over the matching branch.

Covering the amplitude and frequency windows, with Ramsey and echo measurements for low carriers, gives the fixed-error time
\[
 O_\epsilon\!\left(
 \frac{\sqrt{r\log(e+r)}}{\beta_{\min}}+
 \frac{1}{\sqrt{\beta_{\min}\omega_{\min}}}
 \right).
\]
The second term is necessary for bands that extend sufficiently close to zero. Appendix~\ref{app:continuous} gives the pulse, the full-band proof and the corresponding lower bound. The logarithmic factor accounts for the pulse duration and bin spacing needed to control errors through the coherent sequence.

The interrogation-time result does not make the frequency bank or its readout free. A wall-clock accounting has the form
\begin{equation}
T_{\rm wall}=T_{\rm cal}+T_{\rm classical}+
N_{\rm sh}\sum_b(T_{{\rm prep},b}+T_{{\rm int},b}+T_{{\rm read},b}).
\label{eq:wall}
\end{equation}
In particular, fabricating $O(r)$ frequency modes, distributing the couplings $g_k$, providing the global phase reference, compensating static shifts during bus pulses, and maintaining GHZ coherence over each sequence all contribute to physical complexity. Allen \emph{et al.}~\cite{allen} give separate limits for protocols that require large local amplitude QFI at zero field for every carrier, for protocols with bounded coherent memory time, and for protocols restricted to classical modulation of the sensors. The receiver considered here tests a finite displacement and preserves coherence across the auxiliary frequency modes for a branch whose duration grows with $r$. A comparison of hardware cost or noise tolerance would require a common device model.

\section{Numerical simulations}
Figure~\ref{fig:figure2} shows the first-crossing times of the original fixed-sign mark--bus sequence at signal phases $\{0,\pi/2\}$ on sampled carrier grids. We compare the selected code-space and transmon/cQED points with the predicted scale $X=\sqrt{|\Del|}/(mB)^{3/2}$ including the lookup-density factor. The model-specific normalizations preclude a direct comparison of hardware speeds. The Floquet/Sambe points provide a third, truncated model check.

For each sampled parameter set, the first crossing is the smallest tested $Q$ for which $p_{xy}=1-(1-p_x)(1-p_y)\ge1/2$ in two independently prepared quadrature experiments. The code-space and cQED times summarize these individual crossings. Each Sambe star gives the first common crossing of its six sampled cases. Phase averages remain substantial at the tested sizes. Appendix~\ref{app:amplification} derives the large-bandwidth phase dependence for the fixed-sign sequence on a regular bank. Appendix~\ref{app:numerics} gives the sampling, selection and fit details.

\section{Conclusion}
Our analog construction builds on the IQFI description of broadband sensing to express the Grover-like limit in metrological terms. Although the field is weak, the detection task compares the no-signal state with a finite displacement and is not limited by local amplitude estimation at the null. An independent bound makes the sensing-time converse valid for general coherent control without a discretization step.

The multiresonant receiver shows how frequency-selective marks can be combined with coherent mixing rather than explicit signal digitization. We simulated the fixed-sign sequence at two signal phases. A closely related alternating-sign sequence has a proved uniform gap in the stationary reversible-mark model. For a continuously running field, the phase-insensitive sensing pulse of Ref.~\cite{allen} supplies a mark and its inverse to sufficient accuracy for the same bus readout. The resulting uniform detection bound leaves a factor $\sqrt{\log(e+r)}$ in the broadband time. Implementing the bucket-controlled logical rotations with a given device's bandwidth and maintaining coherence throughout the sequence remain experimental questions.

\begin{acknowledgments}
This work was supported by NSERC--NSF alliance grant ALLRP-586858-2023 and an NSERC Discovery grant.
\end{acknowledgments}
\appendix
\section{A finite-displacement bound for general coherent control}\label{app:converse}
Our first derivation of the sensing-time scaling uses the two IQFI regimes with constants uniform along the amplitude path, as displayed in Eq.~\eqref{eq:conditional}. To see the crossover explicitly, for $mBT\ge1$ let $B_c=(mT)^{-1}$. Then Eq.~\eqref{eq:geometry} gives
\begin{equation}
\begin{split}
\KFD(B,T)&\le\frac{C_pm^2TB_c}{B}+\frac{C_nm^3T^2}{B}\int_{B_c}^BB'\dd B'\\
&\le\frac{C_pm}{B}+\frac{C_n}{2}m^3BT^2.
\end{split}
\end{equation}
When $mBT\le1$, using only the perturbative branch gives $\KFD\le C_pm^2T$. A different fixed order-one crossover changes the constants, not the scaling. The argument is conditional on uniform IQFI estimates and on a single fixed control schedule throughout the $B'$ integral.

Here we prove Eq.~\eqref{eq:general} without assuming such a nonzero-amplitude local-IQFI ceiling. Let $\beta=mB$ and work in the interaction picture of the null control. Purify the fixed initial state to a normalized vector $v$ in an extended auxiliary/reference space. The signal propagator satisfies
\begin{equation}
\begin{split}
\ii\partial_tV_{\beta,\omega}(t)&=\beta\cos(\omega t+\phi)\times A(t)V_{\beta,\omega}(t),\\
A(t)&=A(t)^\dagger,\quad\norm{A(t)}\le1.
\end{split}
\end{equation}
Consider a segment $[a,a+\tau]$, with the same input $v$ for every $\omega$. Write $V(a+t,a)v=v+d_\omega(t)$ and
\begin{equation}
F_\omega(t)=\int_a^{a+t}\dd s\,\cos(\omega s+\phi)A(s)v.
\end{equation}
Vector-valued Plancherel, treating the two exponential terms of the cosine separately, gives
\begin{equation}
\norm{F_\omega(t)}_{L^2(\mathbb R,\dd\omega;\mathcal H)}\le\sqrt{2\pi t}.
\end{equation}
The integral equation for $d_\omega$ and Minkowski's inequality imply
\begin{equation}
D(t):=\norm{d_\omega(t)}_{L^2}\le\beta\sqrt{2\pi t}+\beta\int_0^tD(s)\dd s
\le\beta\sqrt{2\pi t}\,e^{\beta t}.
\end{equation}
Thus for an arbitrary segment of length $\tau$,
\begin{equation}
\int_{\mathbb R}\dd\omega\,\norm{[V(a+\tau,a)-\id]v}^2\le2\pi\beta^2\tau e^{2\beta\tau}.
\label{eq:segment}
\end{equation}
One can first restrict the frequency integral to a finite interval and pass to $\mathbb R$; no smoothness or derivative bound on $A(t)$ is needed.

Partition $[0,T]$ into $N=\max\{1,\lceil\beta T\rceil\}$ equal segments $V_1,\ldots,V_N$, so that $\beta\tau\le1$. The exact identity
\begin{equation}
(V_N\cdots V_1-\id)v=\sum_{k=1}^N V_N\cdots V_{k+1}(V_k-\id)v
\end{equation}
has the same $\omega$-independent input $v$ in each difference. Since the preceding factors are unitary at each $\omega$, Cauchy--Schwarz and Eq.~\eqref{eq:segment} yield
\begin{equation}
\begin{split}
\int_{\mathbb R}\dd\omega\,&\norm{[V_{\beta,\omega}(T)-\id]v}^2\\
&\le2\pi e^2\beta^2NT\\
&\le2\pi e^2\beta^2T(1+\beta T).
\end{split}
\label{eq:telescope}
\end{equation}
For normalized purifications $v,w$, $1-|\braket vw|^2\le\norm{v-w}^2$ and $[\arcsin\sqrt{p}]^2\le(\pi^2/4)p$. Fidelity cannot decrease on discarding subsystems. It follows that
\begin{equation}
\int_{\Del}\theta_\omega(B,T)^2\dd\omega\le\frac{\pi^3e^2}{2}\beta^2T(1+\beta T),
\end{equation}
which gives Eq.~\eqref{eq:general} after multiplying by $4/B^2$. The auxiliary dimension, switching rate and strength of $G(t)$ never appear: that control is removed exactly rather than Trotterized. The same Fourier argument at the null gives $K(0,T)\le8\pi m^2T$; the finite-displacement proof does not assert the same bound on local IQFI at arbitrary $B>0$. The theorem assumes a fixed worst-case duration, not an unbounded stopping time with only a finite expectation.

\section{Frequency-bank coverage}\label{app:bank}
Divide the band into $N_0=\lceil r\rceil$ equal cells of width at most $\beta_{\min}$. Draw $N_s=sN_0$ independent, uniformly distributed bucket frequencies with equal bus weights $|g_k|^2=1/N_s$. For the count $C_\ell$ in each cell, the Chernoff bound and a union bound \cite{chernoff} give
\begin{equation}
\Pr\{\exists\ell:|C_\ell-s|>s/2\}\le 2N_0e^{-s/12}.
\end{equation}
Thus $s=O[\log(r/\eta)]$ suffices for every cell to carry weight $\Theta(1/N_0)$ with probability at least $1-\eta$. Counting complete and intersected cells in a detuning window of radius $\beta_j$ proves the uniform coverage estimate, Eq.~\eqref{eq:coverage}, including carriers at the band edges. It is a bank-coverage statement, not a certificate of the full detection probability.

For the alternating-mark theorem choose instead equally weighted midpoints of a regular grid with $N_s=\lceil8r\rceil$. Every interval $|\omega-\nu_k|\le\beta_j/2$ has weight $\Omega(1/r_j)$, where $r_j=|\Del|/\beta_j$. When $r_j$ exceeds a fixed constant, the complementary region $|\omega-\nu_k|\ge4\beta_j$ has weight $\Omega(1)$. Both statements hold uniformly up to either band edge. An iid bank with $\lceil8r\rceil$ equal cells and $s\ge12\log(2\lceil8r\rceil/\eta)$ modes per cell on average has the same properties on the corresponding good-count event, at the expense of $O[r\log(r/\eta)]$ modes.

\section{A continuous-parameter measurement certificate}\label{app:cert}
A finite carrier grid is not a uniform detection certificate unless interpolation and numerical errors are controlled. Consider a predetermined branch $b$ of duration $T_b$, starting at global time $t_b$, whose signal Hamiltonian has the parameter dependence of Eq.~\eqref{eq:hamiltonian}. Duhamel's formula gives
\begin{align}
\norm{\partial_B U_b}&\le mT_b,\qquad\norm{\partial_\phi U_b}\le mB_{\max}T_b,\\
\norm{\partial_\omega U_b}&\le mB_{\max}(t_bT_b+T_b^2/2).
\end{align}
A corresponding fixed-measurement rejection probability is Lipschitz with constants $L_B=2mT_b$, $L_\phi=2mB_{\max}T_b$, and $L_\omega=mB_{\max}(2t_bT_b+T_b^2)$. Use a product grid of spacings $h_B,h_\omega,h_\phi$, including both amplitude and frequency endpoints and wrapping periodically in phase. A sufficient margin for branch $b$ is
\begin{equation}
\mu_b=\tfrac12(L_Bh_B+L_\omega h_\omega+L_\phi h_\phi)+\varepsilon_{\rm num}+\varepsilon_{\rm model}.
\label{eq:mesh}
\end{equation}
Take $\mu=\max_b\mu_b$ across the predetermined branches. If the minimum over sampled parameter points of $\max_b p_b$ is at least $p_1+\mu$, and the aggregate null ceiling obeys $\sum_b p_b(0)\le p_0$, Eq.~\eqref{eq:family} follows. For experimentally sampled probabilities, simultaneous binomial error margins must be added to $\mu$ and to the null budget, with their own calibration-failure probability. These derivative bounds are sufficient, not a claim that the required calibration has already been performed. They also do not cover extra parameter-dependent implementation terms unless those terms are bounded separately.

\section{Stationary marks, unknown phase, and the two-reflection receiver}\label{app:amplification}
\subsection{Uniform detection from alternating signs}
For the centered rectangular mark of duration $2\pi/\beta_j$ and $\pi$ mixer, the stationary scalar-mark assumption makes the GHZ eigenbranch phase
\begin{equation}
\begin{split}
\eta_k&=\pi x\cos\alpha\;\sinc\!\left(\pi\frac{\omega-\nu_k}{\beta_j}\right),\\
x&=\frac{B}{B_j}\in[1,2),\quad\sinc z=\frac{\sin z}{z}.
\end{split}
\label{eq:eta}
\end{equation}
The relative reference angle $\alpha$ is drawn uniformly, independently of the unknown signal, and remains fixed through a branch. A known $\pi$ reference shift is assumed to make alternating marks $D$ and $D^\dagger$ exactly. On one GHZ branch, two mark--bus cycles act as $W=RD^\dagger RD$; since $D^\dagger RD=R_{D^\dagger u}$, $W$ is a product of reflections. In their common two-dimensional invariant plane, $\cos\vartheta=|\braket u{Du}|$ and
\begin{equation}
\braket u{W^q u}=\cos(2q\vartheta).
\end{equation}
The other GHZ eigenbranch has the same real return. For the calibrated complete-input projector, the rejection probability is therefore $p_q=\sin^2(2q\vartheta)$, exactly.

Fix an allowed $(B,\omega,\phi)$ and choose the matching amplitude window, so $x=B/B_j\in[1,2)$. Restrict the random reference to $c_-\le|\cos\alpha|\le c_+$, where $0<c_-<c_+<1/6$ and $c_+<4c_-$. This event has a probability $g>0$ independent of the signal parameters. On this event, $|\eta_k|<\pi/3$, the near-bucket phases have magnitude at least $2c_-$, and the far-bucket phases have magnitude at most $c_+/2$. The separation $2c_--c_+/2$ is strictly positive.

Write $w_k=|g_k|^2$. The identity
\begin{equation}
\sin^2\vartheta
=1-\left|\sum_kw_ke^{-\ii\eta_k}\right|^2
=\sum_{k,\ell}w_kw_\ell[1-\cos(\eta_k-\eta_\ell)]
\label{eq:anglebound}
\end{equation}
then combines the near and far weights from Appendix~\ref{app:bank} to give $\sin^2\vartheta\ge a/r_j$ for some absolute $a>0$, provided $r_j$ exceeds a fixed constant. The same phase restriction ensures $\cos\vartheta\ge c>0$, so $|\sin(2\vartheta)|\ge c'\!/\sqrt{r_j}$. If $q$ is uniform in $\{1,\ldots,L_j\}$ with $L_j=\lceil C\sqrt{r_j}\rceil$ for a sufficiently large fixed $C$, the finite geometric sum implies
\begin{equation}
\mathbb E_q\sin^2(2q\vartheta)
\ge \frac12-\frac{1}{2L_j|\sin(2\vartheta)|}
\ge\frac14.
\end{equation}
For the remaining bounded range of $r_j$, apply one mark and test the complete input. The two GHZ branches give return amplitude $\sum_k w_k\cos\eta_k$. On the same reference event all cosines are positive, the near buckets have order-one weight, and their cosines are at most $\cos(2c_-)<1$. The rejection probability is thus bounded below by a positive constant. Thus the predetermined dyadic scan has $p_{\rm null}=0$ and $\inf_{B,\omega,\phi}p_{\rm signal}>0$, with constants independent of $r$, $B$, $\omega$, $\phi$ and $m$. The required number of mark pairs is $O(\sqrt{r_j})$. The same reasoning applies on the good-count event of an iid bank, with modified absolute constants.

\subsection{Fixed-sign marks}
The original simulations repeatedly apply the same-sign centered $2\pi$ mark $D$ and reflection $R$, computing $(RD)^Q$ for the represented carrier/phase samples. For a general bucket kick $R_\vartheta=\id+(e^{-\ii\vartheta}-1)\ket u\bra u$, the secular equation for an eigenvalue $e^{-\ii\lambda}$ of $U=R_\vartheta D$ away from the poles is
\begin{equation}
\begin{split}
\sum_k w_k\cot\!\left(\frac{\eta_k-\lambda}{2}\right)&=-\cot\frac{\vartheta}{2},\\
\Gamma(\lambda)&=\sum_k w_k\csc^2\!\left(\frac{\eta_k-\lambda}{2}\right).
\end{split}
\label{eq:green}
\end{equation}
It runs over all buckets coupled to $\ket u$. Restricting to the matched sector and renormalizing its weights changes the physical mixer; a local Green-function estimate does not by itself give an overlap or detection guarantee for the full receiver. Similarly, a local simple-crossing Taylor expansion must not be applied at a quadratic resonance without checking its hypothesis.

In the fixed-sign simulations the mark duration is matched to the signal amplitude, $\tau_M=2\pi/(mB)$, so $x=1$ in Eq.~\eqref{eq:eta}. At a relative phase $\pi/4$, both quadratures then obey $|\eta_k|\le\pi/\sqrt{2}$, uniformly separated from the reflection resonance at $\pi$. The full-bank secular equation gives a mode with eigenphase $\pi+\varepsilon$, initial-state weight $v_0=1/\Gamma_\varepsilon$, and
\begin{equation}
\begin{split}
&\sum_kw_k\tan\!\left(\frac{\eta_k-\varepsilon}{2}\right)=0,\\
&\Gamma_\varepsilon=\sum_kw_k\sec^2\!\left(\frac{\eta_k-\varepsilon}{2}\right).
\end{split}
\end{equation}
The remaining spectral weights sum to $1-v_0$. For independent quadrature measurements, $p_{xy}=1-(1-p_x)(1-p_y)$. Retaining both GHZ sensor branches gives
\begin{equation}
p_{xy}(Q)\le\min\{1,2\sin^2(Q\varepsilon)+8(1-v_0)\}.
\end{equation}
On a flat regular bank the sinc tail gives $\sum_k w_k|\eta_k|=O(\log r/r)$ and $\sum_k w_k\eta_k^2=O(1/r)$. The tangent sum at $\varepsilon=0$ is therefore $O(\log r/r)$. Its derivative near zero is at most $-1/2$, with no pole there, so it has a root $\varepsilon=O(\log r/r)$. Then $\Gamma_\varepsilon-1=\sum_k w_k\tan^2[(\eta_k-\varepsilon)/2]=O(1/r)$, giving $1-v_0=O(1/r)$. Consequently $p_{xy}\to0$ at $Q=O(\sqrt{r\log r})$ for this phase. At each relative phase off the quadrature axes, both quadratures retain a positive separation from resonance. The same estimates, with phase-dependent constants, give pointwise convergence to zero. Since $0\le p_{xy}\le1$, dominated convergence gives a vanishing uniform-phase average for this regular bank at the same schedule. This does not contradict the recorded finite-size crossings at the two sampled phases $0,\pi/2$; it shows why neither those crossings nor local frequency-bank coverage can prove phase-uniform performance. The alternating-sign receiver has different dynamics and is covered by the preceding theorem.

\section{Physical flag model and sequence timing}\label{app:physics}
An idealized circuit-QED conditional-level-shift model \cite{blais} uses physical sensor/flag states $\ket g_a,\ket e_a$ (not the logical GHZ branches) and
\begin{equation}
\begin{split}
H_0&=(A_{\rm aux}\oplus0_c)\otimes\sum_{a=1}^m\ket e_a\bra e_a,\\
H_{\rm sig}&=Bz_j(t)\cos(\omega t+\phi)\times\left(\sum_kP_k\right)\otimes\sum_a X_a.
\end{split}
\end{equation}
Here $z_j(t)$ gates the mark window and $X_a$ acts on the two physical levels of sensor $a$. The projector $\sum_kP_k$ assumes that the signal coupling is inactive in the bus sector. In the interaction picture of $\nu_k\ket e\bra e$, a transverse physical drive generally produces both $X$ and $Y$ quadratures, rather than a scalar cosine multiplying a fixed $J_\chi$. Such dynamics need not preserve the two-branch GHZ code for $m>1$. The effective commuting mark in Eq.~\eqref{eq:markham} must therefore be supplied by additional control, or its replacement analyzed in the full physical register.

One explicit route to a single scalar mark is bucket-conditioned product modulation,
\begin{equation}
H^{\rm prod}_{k,\chi}(t)=B f_M(t/\tau_M)\cos(\omega t+\phi)\cos(\nu_kt+\phi_\chi)\sum_a Q_\chi^{(a)}.
\label{eq:product}
\end{equation}
Here $Q_\chi^{(a)}$ is the Pauli operator along $\chi$ on sensor $a$. The cosine-product identity gives a slow difference-frequency term and a fast sum-frequency term. In the two-level restriction, all these signal operators commute within one mark window; their integral gives the GHZ mark with a calculable counterrotating correction. Realizing this product modulation requires actual conditional control: it cannot be obtained merely by changing the frame of a generic transverse drive.

More importantly, the absolute start time $t_n$ in Eq.~\eqref{eq:mark} cannot be erased. A known bucket-dependent phase can compensate a known $\nu_k t_n$ offset, but not the unknown $\omega t_n$. Consequently the identical marks used in the fixed-sign simulations and the exact inverses used in the alternating-sign proof are stationary sequence assumptions, not consequences of a single-pulse reduction. A shift of $\pi$ reverses a scalar mark at the same relative phase; after a later unknown detuning advance the magnitude generally changes. The continuously driven construction in Appendix~\ref{app:continuous} instead bounds the error for a smooth sensing pulse uniformly in its starting phase.

During the bus pulse, the static sensor-dependent shifts can also make the interaction-picture bucket--bus exchange sensor dependent. The star mixer in Eq.~\eqref{eq:bus} requires compensation, echoing, or resolved control. An eliminated detuned bus produces an effective rank-one phase $-|\lambda|^2\ket u\bra u/\Delta_c$ only within a controlled dispersive regime, and its time is $O(|\Delta_c|/|\lambda|^2)$.

\section{Numerical methods and interpretation}\label{app:numerics}
Figure~\ref{fig:figure2} reports three distinct numerical models. The GHZ code-space calculation applies the commuting scalar mark and star reflection in dimension $2(N_s+1)$. Dense-code and full-tensor checks agree with this reduction at numerical precision for the tested instances. The overlays use the Floquet formalism of Sambe~\cite{sambe}, retaining three harmonics of the transverse drive. Sideband convergence has not been established. The effective transmon/cQED model uses three Duffing levels per sensor, a scalar-demodulated mark, first-order Magnus propagation and a star bus. Neither set of overlays is a noise-aware device simulation.

All three models sample phases $0$ and $\pi/2$. The 70 code-space groups use $m=1$--$5$ and $r=2^5,\ldots,2^{18}$, with a regular bank extended by $4mB$ at each edge. In ascending order of $r$, the numbers of modes per linewidth are $s=4,4,4,4,6,6,8,12,14,16,16,20,22,24$, giving $N_s=s(r+8)$. Five seeded grids per group each contain five carriers. The time for a grid is the maximum of its ten individual first crossings. The point and whiskers give the median and range of the five grid times. Of 1750 carrier draws, 50 fall outside the nominal band in the guard region.

The 34 Sambe groups use $m=1$--$4$, $r=128$--$131072$, and $N_s=528$--$2883672$ with three off-grid carriers per group. The 17 selected transmon/cQED groups use $m=1$--$5$, $r=64$--$65536$, and $N_s=244$--$2097152$. Each transmon group uses a single sampled bank and seven carriers from the central half of the band, except for nine carriers at $(m,r)=(1,65536)$. Its point and upper whisker give the median and maximum individual first-crossing times. These samples provide no estimate of the random-bank failure probability $\eta$.

The group-level fits use each group's actual band ratio $r$ in $T=AX\sqrt{q_{\log}(r)}$, with $X=\sqrt{|\Del|}/(mB)^{3/2}$. The separately fitted density schedules are
\begin{equation}
\begin{split}
q_{\log}^{\rm code}(r)&=\max\{4,-7.881+1.679\log_2r\},\\
q_{\log}^{\rm cQED}(r)&=\max\{3.813,-13.588+2.873\log_2r\}.
\end{split}
\end{equation}
These fits give RMS log residuals of $0.169$ (code) and $0.155$ (cQED). To draw one curve per population against $X$, Fig.~\ref{fig:figure2} uses $B=1$ and $r=\max\{2,(m_{\rm ref}X)^2\}$, fitting $m_{\rm ref}\in[1,4]$ and $A$. The plotted curves have $m_{\rm ref}=1.61$ and $1.00$, with RMS log residuals of $0.20$ and $0.19$, respectively. Fitting $T=A_\alpha[X\sqrt{q_{\log}(r)}]^\alpha$ by unweighted least squares in $\log T$ gives $\alpha=1.072$ and $1.004$, respectively. To assess sensitivity to the represented ratios, we resampled whole $r$ blocks 2000 times, refitting the density schedule in each resample. The central 95 percent of the fitted slopes lay in $[1.041,1.087]$ and $[0.941,1.092]$, respectively. These ranges describe variation within the selected data. The reported cQED rows were filtered for resolved crossings and agreement with the ideal model, and some higher-ratio code-space source runs were unresolved. The selected groups cannot establish a failure probability or a uniform phase-and-frequency guarantee. The cQED inset uses separately selected banks at $(m,r)=(5,64),(4,128),(3,256),(2,32768)$. For example, at $(5,64)$ the 1536-mode inset bank gives $0.509$, while the 244-mode timing bank gives $0.154$. The fixed-sign limitation is analyzed in Appendix~\ref{app:amplification}.

\section{Accumulated error and resource budgets}\label{app:error}
For a branch of $Q$ ideal cycles, if the implemented mark and mixer unitaries have per-cycle operator errors at most $\epsilon_M$ and $\epsilon_S$, the telescoping identity gives
\begin{align}
\norm{\widetilde U-U}\le Q(\epsilon_M+\epsilon_S),\\
|\widetilde p-p|\le2Q(\epsilon_M+\epsilon_S)
\end{align}
for a fixed pure-input measurement probability. With open-system noise, a per-cycle error at most $\epsilon_{\rm cyc}$ in half the diamond norm bounds the final probability error by $Q\epsilon_{\rm cyc}$, plus preparation and measurement errors. Summing branch budgets bounds the complete trial. A sufficient (conservative) condition to preserve a gap $\Delta p$ is that the accumulated probability error under each hypothesis be less than $\Delta p/4$. In particular, a noise-rate budget $\int_0^{T_b}\gamma(t)\dd t\ll\Delta p$ requires device-specific estimates of relaxation, dephasing, resonator loss, disorder, pulse distortion, thermal occupation and crosstalk. These are bounds on tolerated imperfections, not evidence that any particular implementation meets them.

\section{Detection with a continuously running field}\label{app:continuous}

We give a choice of sensing pulses for which the field in Eq.~\eqref{eq:hamiltonian} may run throughout the experiment. The pulses are the elementary sensing units of Allen \emph{et al.}~\cite{allen}. We use the phase they produce directly in the two-reflection receiver of Eq.~\eqref{eq:reflections}.

\paragraph*{One frequency band and amplitude window.}
Let the frequency band be $[a,a+w]$ with $a\ge w>0$, and let $B_0\le B\le 2B_0$ as in the dyadic amplitude scan. Write $b_0=mB_0$, $b=mB$, and $x=w/b_0$. For sufficiently large $x$, there is a test with zero null rejection and a positive constant rejection probability for every such $B,\omega,\phi$, in time
\begin{equation}
 T_{\rm win}=O\!\left(\frac{\sqrt{x\log x}}{b_0}\right).
 \label{eq:continuous-win-time}
\end{equation}
We take the sensor generators to be Pauli $Z$ and allow arbitrarily fast, bucket-controlled logical sensor rotations. On the code spanned by $\ket{0_L}=\ket{0}^{\otimes m}$ and $\ket{1_L}=\ket{1}^{\otimes m}$, the field generator is exactly $mZ_L$. A logical rotation preserves this code and can require joint control of the $m$ sensors. The bucket register uses $N=\Theta(x/\log x)$ modes and one bus.

\paragraph*{The sensing pulse.}
Choose $N=\lfloor x/(4A\log x)\rfloor$, with $A$ a sufficiently large fixed constant, and set $h=w/(4N)$. Thus $h\sim A b_0\log x$. Following Ref.~\cite{allen}, partition the band into $N$ groups of four adjacent intervals of width $h$. One bank covers the first and third intervals of each group, with its reference frequency $3h/2$ above the group's lower edge. The other covers the second and fourth with reference $5h/2$ above that edge. For each $\omega$ there is a bank and an index $k_*$ with $h/2\le|\omega-\nu_{k_*}|\le3h/2$ and $|\omega-\nu_k|\ge5h/2$ for $k\ne k_*$. This choice keeps all detunings in that bank away from zero, including at the band edges.

Let $t_0$ denote the absolute start of a pulse and $s$ the time from its start. Lemma S2 of the Supplemental Material of Ref.~\cite{allen} gives controls in bucket $k$ that realize the Hamiltonian
\begin{align}
 H_k(s)&=b f(s)\cos(\omega s+\phi+\omega t_0)\nonumber\\[-2pt]
 &\hspace{1em}\times [X_L\cos(\nu_k s)-Y_L\sin(\nu_k s)]
 \label{eq:continuous-pulse}
\end{align}
to arbitrarily small propagator error. Here $f(s)=F(s/\tau)$ with $F(v)=\exp[-1/(v(1-v))]$ on $0<v<1$ and zero elsewhere. Set $I_2=\int_0^1 F(v)^2\,\dd v$, $C_T=\pi/(8I_2)$, and $\tau=C_T h/b_0^2$. Write $\Delta_k=|\omega-\nu_k|$. On the successful bank $h/2\le\Delta_k\le w\le\omega$. Proposition S1 in the Supplemental Material of Ref.~\cite{allen} bounds the rotating-wave error by $C(b/\omega)(1+b\tau)$ uniformly in the starting phase.

In the detuning frame the rotating-wave equation is $\ii\partial_v\psi=(\tau\Delta_k)H(v)\psi$, where $v=s/\tau$, $H(v)=(\pm Z_L+\lambda_k F(v)X_{\xi})/2$, $\lambda_k=b/\Delta_k\le 4b_0/h$, and $X_{\xi}$ is a transverse Pauli operator of fixed azimuth. Its gap is at least one. The pulse is flat at both endpoints. After subtracting the instantaneous ground energy, the normalized Hamiltonian has uniform Gevrey-2 derivative bounds $\|\partial_v^n H\|\le K_0 K_1^n(n!)^2$ for $n\ge1$ and $|\lambda_k|\le1$. These bounds hold for $F$ and for analytic functions of $\lambda_k F$ on this range. Choose fixed $K_0,K_1\ge1$ uniformly in $\lambda_k$, increasing $K_1$ to absorb the factor $(n+1)^{-2}$ in the Gevrey bound. Supplemental Theorem~1 of Ge \emph{et al.}~\cite{ge_adiabatic} then bounds the final transition amplitude by $C\exp[-c\sqrt{\tau\Delta_k}]$. Changing the start phase conjugates the rotating-wave evolution by a $Z_L$ rotation. Its diagonal entries therefore give the same return phase $\theta_k$ for every start. Unitarity and unit determinant give conjugate return amplitudes on the two sensor levels.

The angle in the mark includes the correction to the dynamical phase. With the free phase removed, the latter has magnitude
\[
 a_k=\frac{\tau\Delta_k}{2}\int_0^1
 [\sqrt{1+\lambda_k^2F(v)^2}-1]\,\dd v .
\]
The instantaneous eigenvector in parallel-transport gauge has length of path at most $\tfrac12|\lambda_k|\int|F'|$, from its polar angle $\arctan(\lambda_k F)$. Integrating the derivative of the backwards-propagated eigenvector after removing its eigenenergy gives $\big||\theta_k|-a_k\big|=O(|\lambda_k|)$ on the small-angle branch. For $|z|\le1$, $z^2/3\le\sqrt{1+z^2}-1\le z^2/2$. All angles on the successful bank obey $|\theta_k|\le\pi/4+o(1)$. At the same signal amplitude $b$, the matched and other detunings give, for $k\ne k_*$,
\[
 |\theta_{k_*}|-|\theta_k|
 \ge \frac{b^2\tau I_2}{h}\left(\frac19-\frac1{10}\right)-o(1).
\]
Since $b^2\tau/h\ge C_T$, the separation is bounded below by a positive constant. Hence for some $d>0$,
\begin{align}
 \cos\theta_{k_*}&\ge0, &
 \cos\theta_k-\cos\theta_{k_*}&\ge d\quad(k\ne k_*).
 \label{eq:continuous-contrast}
\end{align}

Let $V_*=\sum_kP_k\otimes e^{-\ii\theta_k Z_L}$ and let $V(t_0)$ denote a pulse at absolute time $t_0$ with the known control frame closed. The direct sum over buckets takes the maximum error. For $h\sim A b_0\log x$ with sufficiently large fixed $A$, the transition error is $O(x^{-2})$ and the rotating-wave error is $O(\log x/x)$. Thus $\|V(t_0)-V_*\|\le e_x$ uniformly in $t_0$, with $\sqrt N e_x\to0$. The modulation and conditional rotations in Lemma S2 of Ref.~\cite{allen} can be synthesized to $o(N^{-1/2})$ additional error per mark. In replacing instantaneous controls by fast pulses with the same frame endpoints, the two signal generators differ by at most $2mB_{\max}$ during the replacement intervals and coincide elsewhere. Their total length can be chosen correspondingly small. At zero field the known controls close exactly to the identity.

\paragraph*{Reading out the accumulated phase.}
Prepare $\ket u=N^{-1/2}\sum_k\ket k$ and $\ket{0_L}$. On this sensor branch $V_*$ acts as $D=\sum_k e^{-\ii\theta_k}P_k$. Since $X_L Z_L X_L=-Z_L$, the operation $X_L V(t_1)X_L$ in a later window approximates $V_*^\dagger$ with the same error, for any $t_1$. Use these two types of pulses alternately, with the $\pi$ bus reflection $R$ after each. The field acting during a bus pulse commutes with its sensor-independent control and contributes a common phase on the ideal sensor branch. The bus pulse may have duration $O(1/b_0)$.

Put $z=N^{-1}\sum_k e^{-\ii\theta_k}$ and $\vartheta=\arccos|z|$. Equation~\eqref{eq:continuous-contrast} implies $|z|\ge\operatorname{Re}z\ge (N-1)d/N$ and
\begin{equation}
 1-|z|^2=\frac{1}{N^2}\sum_{k<l}|e^{-\ii\theta_k}-e^{-\ii\theta_l}|^2
 \ge\frac{(N-1)d^2}{N^2}.
\end{equation}
The last inequality retains the $N-1$ pairs containing $k_*$. Hence $\sin(2\vartheta)\ge c/\sqrt N$ for some $c>0$. Under the ideal pair $RD^\dagger RD$, the probability of rejecting $\ket u$ after $q$ pairs is $\sin^2(2q\vartheta)$. Its average over $q$ uniform on $1,\ldots,Q$ is at least $1/2-1/[2Q\sin(2\vartheta)]$, by the finite geometric sum. Taking $Q=\lceil2\sqrt{N}/c\rceil$ gives rejection at least $1/4$. For the exact sensor-independent bus, the $2q$ physical pulses change the final state vector by at most $2Qe_x=o(1)$, after including the common bus phases in the ideal evolution. A mixer error $e_R$ per call adds at most $2Qe_R$ to this bound (Appendix~\ref{app:error}). The physical rejection probability therefore retains a constant gap when $Qe_R$ is sufficiently small. At the null every mark is the identity and $R^2=\id$. Running both banks on fresh probes and rejecting if either rejects gives the claimed uniform gap. The maximum time is $O(\sqrt N(\tau+1/b_0))$, which gives Eq.~\eqref{eq:continuous-win-time}.

\paragraph*{Covering the full signal promise.}
Write $W=|\Del|$, $\beta_{\min}=mB_{\min}$ and $r=W/\beta_{\min}$. Assume $mB_{\max}\le W$, as in the main text. With the same ideal coherent controls, the test has zero false-alarm probability and misses any promised signal with probability at most a fixed $\epsilon>0$ in total elapsed time
\begin{equation}
 T_{\rm cont}=O_\epsilon\!\left(
 \frac{\sqrt{r\log(e+r)}}{\beta_{\min}}
 +\frac{1}{\sqrt{\beta_{\min}\omega_{\min}}}\right).
 \label{eq:continuous-full-time}
\end{equation}
Elapsed interrogation time is measured from a common start and includes any waits. A tone can start at a zero crossing when the lower band edge approaches zero.

Use the dyadic amplitude windows of Sec.~IV, with $b_0=2^j\beta_{\min}$. Fix a constant $K\ge1$ large enough for Eq.~\eqref{eq:continuous-win-time} to hold whenever $x\ge K$. In one amplitude window, set $A=\max(\omega_{\min},Kb_0)$. To cover $[A,\omega_{\max}]$ when nonempty, let $V=\omega_{\max}-A$. If $V\le A$, use one band starting at $A$ of width $\max(Kb_0,V)$. If $V>A$, use successive bands $[2^jA,2^{j+1}A]$ until the upper edge is covered. Each band has width between $Kb_0$ and its lower carrier. In the second case the widths grow geometrically and the largest is at most $2W$. In the first case the width is at most $KW$, since $b_0\le W$. Summing Eq.~\eqref{eq:continuous-win-time} thus costs $O(\sqrt{(W/b_0)\log(e+W/b_0)}/b_0)$.

Below $Kb_0$, partition the remaining carriers into intervals $[d,2d]$ with $d=2^j\omega_{\min}\le\min(\omega_{\max},Kb_0)$. On each interval use two consecutive fresh GHZ probes in $(\ket{0_L}+\ket{1_L})/\sqrt 2$. Give the first a Ramsey interrogation of length $t_R=c_R/b_0$. Give the second a Hahn echo of length $t_E=c_E/\sqrt{b_0d}$ with a logical $X_L$ pulse at its midpoint. Measure each probe in the logical $X_L$ basis. If $\alpha_R$ and $\alpha_E$ are the field phases at the midpoints of the two shots, the rejection probabilities are $\sin^2\theta_R$ and $\sin^2\theta_E$, where direct integration gives
\begin{align}
 \theta_R&=b t_R\,\frac{\sin(\omega t_R/2)}{\omega t_R/2}\cos\alpha_R,\nonumber\\
 \theta_E&=\frac{4b}{\omega}\sin^2(\omega t_E/4)\sin\alpha_E.
 \label{eq:continuous-low-quadratures}
\end{align}
For instance choose $c_R=1/(16K)$ and $c_E=1/(16\sqrt K)$. Then $|\alpha_E-\alpha_R|=\omega(t_R+t_E)/2\le1/8$. The coefficients of the cosine and sine in Eq.~\eqref{eq:continuous-low-quadratures} lie in $[2c_R/\pi,2c_R]$ and $[c_E^2/\pi^2,c_E^2]$, respectively, by $2y/\pi\le\sin y\le y$ for $0\le y\le\pi/2$. Both intervals are inside $(0,\pi/2)$. If $|\cos\alpha_R|<1/2$, then $|\sin\alpha_E|\ge\sqrt3/2-1/8>1/2$. Consequently at least one shot rejects with probability bounded below by a positive constant for every phase. Both shots accept with certainty at zero field. Their duration is $O(1/\sqrt{b_0d})$ since $d\le Kb_0$. Summing over $d$ costs $O(1/\sqrt{b_0\omega_{\min}})$.

Reject if any test rejects. For each signal, some amplitude window and carrier interval has the stated constant detection probability for every start time. Under the null every interval accepts, so the number of intervals carries no false-alarm penalty. A constant number of repetitions at fixed $\epsilon$ gives the desired miss probability. The sums over amplitude windows are geometric in $b_0^{-3/2}$ and $b_0^{-1/2}$, proving Eq.~\eqref{eq:continuous-full-time}.

The dependence on the lower carrier in Eq.~\eqref{eq:continuous-full-time} is necessary when $\omega_{\min}$ is sufficiently small. For $b=\beta_{\min}$ and $\phi=\pi/2$, the trace distance between signal and null outputs after elapsed time $T$ is at most $\beta_{\min}\int_0^T|\sin(\omega_{\min}t)|\,\dd t\le\beta_{\min}\omega_{\min}T^2/2$. This follows by integrating the difference of the propagators in the control frame, and persists under measurement. A constant detection gap therefore requires $T=\Omega(1/\sqrt{\beta_{\min}\omega_{\min}})$.

\bibliography{bib}
\end{document}